%% file: apssamp.tex
\documentclass[
  aps,
  pra,          
  reprint,
  superscriptaddress
]{revtex4-2}

\usepackage{enumitem}
\usepackage[dvipsnames]{xcolor}

\usepackage{amsmath,amssymb}
\newcommand{\ket}[1]{\left|#1\right\rangle}
\newcommand{\bra}[1]{\left\langle#1\right|}
\newcommand{\ketbra}[2]{\ket{#1}\!\bra{#2}}

\usepackage{graphicx}
\usepackage{dcolumn}
\usepackage{bm}

\usepackage{orcidlink}
\usepackage{hyperref} 

\begin{document}

\preprint{APS/123-QED}

\title{Digital-Analog Quantum Computing with Qudits}

\author{Alatz Alvarez-Ahedo
\orcidlink{0000-0002-0081-9759}}
\email{alatz.alvarez@ehu.eus}
\affiliation{Department of Physical Chemistry, University of the Basque Country UPV/EHU, Apartado 644, 48940 Leioa, Spain}
\affiliation{EHU Quantum Center, University of the Basque Country UPV/EHU, Barrio Sarriena s/n, 48940 Leioa, Spain}

\author{Mikel Garcia de Andoin
\orcidlink{0000-0002-5009-7109}}
\affiliation{Department of Physical Chemistry, University of the Basque Country UPV/EHU, Apartado 644, 48940 Leioa, Spain}
\affiliation{EHU Quantum Center, University of the Basque Country UPV/EHU, Barrio Sarriena s/n, 48940 Leioa, Spain}

\author{Mikel Sanz
\orcidlink{0000-0003-1615-9035}}
\affiliation{Department of Physical Chemistry, University of the Basque Country UPV/EHU, Apartado 644, 48940 Leioa, Spain}
\affiliation{EHU Quantum Center, University of the Basque Country UPV/EHU, Barrio Sarriena s/n, 48940 Leioa, Spain}
\affiliation{IKERBASQUE, Basque Foundation for Science, Plaza Euskadi 5, 48009 Bilbao, Spain}
\affiliation{Basque Center for Applied Mathematics (BCAM), Alameda Mazarredo 14, 48009 Bilbao, Spain}

\begin{abstract}
	Digital–analog quantum computing with two-level systems is a computational paradigm that combines an analog Hamiltonian with single-qubit gates to achieve universality. We extend this framework to $d$-level systems by conjugating an analog Hamiltonian block with single-qudit gates drawn from the Weyl–Heisenberg basis, which provides a natural set of operations for qudit architectures. More specifically, we propose a protocol to simulate arbitrary two-body Hamiltonians with at most $\mathcal{O}(d^4 n^2)$ analog blocks. The power of this approach is illustrated by the simulation of many-body qudit spin Hamiltonians including magnetic quadrupolar terms. 
\end{abstract}

\maketitle

\section{Introduction}
 
Quantum computation originated from the idea of simulating quantum systems with controllable quantum simulators \cite{Benioff1980, Feynman1982}. Firstly, it was proposed a paradigm called \textit{analog quantum computing} (AQC) \cite{Das2008}, in which a controllable quantum system, called a quantum simulator, emulates the dynamics of the quantum system of interest. Afterwards, a computational paradigm which proceeds through sequences of discrete quantum gates acting on one or more qubits was proposed, named \textit{digital quantum computing} (DQC) \cite{Deutsch1995}. While AQC is generally more robust to noise \cite{Poggi2020, Altman2021, adrian2024, Fauseweh2024}, it lacks the versatility of DQC, whose gate-based formulation enables universal quantum computation. In contrast, the susceptibility of DQC to noise can be suppressed by employing \textit{quantum error correction codes} \cite{aharonov1996, Kitaev1997, Bultrini2022}. These codes, however, impose substantial resource overheads, as a single logical qubit requires multiple high-quality physical qubits. Although recent progress in hardware development is moving towards larger implementations with better quality qubits \cite{Wintersperger2023,Bluvstein2024, Bravyi2024,strohm2024ionbasedquantumcomputinghardware,AbuGhanem2025}, current platforms still operate within the \textit{noisy intermediate-scale quantum} (NISQ) regime \cite{Preskill2018NISQ}.

To improve the performance of quantum processors in the NISQ era, a hybrid framework combining the robustness of AQC and the flexibility of DQC, known as \textit{digital-analog quantum computing} (DAQC) \cite{Parra2020}, was proposed. DAQC integrates both paradigms by employing the system's natural two-body Hamiltonian as an entangling resource, while digital single-qubit rotations are used to conjugate the Hamiltonian to yield a universal set of operations \cite{PhysRevA.65.040301, Bassler2023synthesisof, ArbitraryDAQC}. This allows for the implementation not only of quantum simulations of arbitrary two-body Hamiltonians, but also quantum algorithms \cite{Ana2020QFT,Ana2022QAOA,Ana2023HHL}. Within the DAQC paradigm, two implementation schemes are typically considered: \textit{stepwise-DAQC} (sDAQC), where the Hamiltonian is switched on and off between single-qubit rotations, and \textit{banged-DAQC} (bDAQC), in which the Hamiltonian remains active while single-qubit gates are applied on top of the continuous dynamics. Although the latter introduces an intrinsic error, it is experimentally more feasible, since the interaction Hamiltonian is not switched on and off, and exhibits superior scalability with respect to the dominant error sources in current quantum processors \cite{Garcia2024}.

Universal quantum computation can also be achieved using qudits through appropriate two-qudit gates or an entangling Hamiltonian and local unitaries \cite{Nielsen2002, Bremner2005, Brennen2005, Brennen2006}. Encoding quantum information in $d$-level quantum systems, known as qudits, enhances the computational capabilities of near-term devices by reducing the number of entangling operations, which might lead to improved experimental implementations \cite{Wang2020, Nikolaeva2024, Kiktengo2025, Gedik2015}. Qudits offer distinct advantages in quantum simulation, where they provide a natural platform for studying high-dimensional spin phenomena \cite{Senko2015, Wang2023}, enabling, for instance, qubit–qudit hybrid algorithms \cite{Bocharov2017, Meth2025}. Qudit-based processors have already been experimentally demonstrated on superconducting platforms \cite{Goss2022, Luo2023, Nguyen2024, Yu2025} and trapped-ion architectures \cite{Chi2022, Ringbauer2022}, among others.

In Ref.~\cite{Nielsen2002}, it was proven that universal quantum simulation can be achieved employing a fixed two-body entangling $N$-qudit Hamiltonian and local unitaries, but they provided no constructive method. Afterwards, in Ref.~\cite{Bremner2005}, they provided a method to achieve it, but with a worse circuit complexity than DAQC.

In this article, we aim to generalize the DAQC protocols \cite{Parra2020,ArbitraryDAQC} to qudit systems. More specifically, we extend the DAQC framework to simulate a target two-body qudit Hamiltonian using another arbitrary two-body qudit Hamiltonian as a resource, together with local qudit unitaries. Our constructive method allows us to design DAQC circuits which employ polynomially fewer applications of single-qudit gates than previous protocols relying on isolating every coupling~\cite{Bremner2005}.

The article is organized as follows: section~\ref{sec:ClassicDAQC} reviews the DAQC paradigm and the implementation of the arbitrary two-body Hamiltonian protocol and discusses the spin-basis case for any $d$; section~\ref{sec:DAQCqudit} presents the extension to $\mathrm{SU}(d)$ systems; section~\ref{sec:DAQCqutritEx} discusses its implementation and performance relative to digital quantum circuits; and section~\ref{sec:Conclusions} summarizes our main results and outlines future directions.

\section{Revisiting the DAQC protocol for qubit Hamiltonians}
\label{sec:ClassicDAQC}

\begin{figure}[t]
	\includegraphics[width=0.48\textwidth]{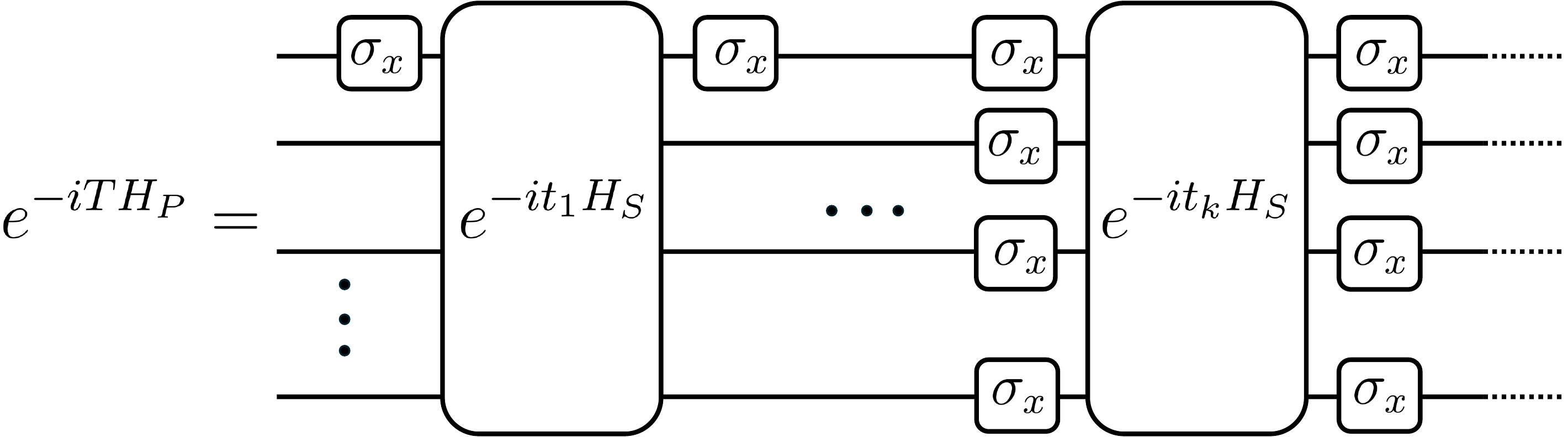}
	\caption{Example of the structure described in Eq.~(\ref{eq:SIS_Exp_Sum}), where $G_1 = \sigma_x^1$ and $G_k = \bigotimes_{i=1}^n \sigma_x^i$ illustrate single-body local operators that conjugate the Hamiltonian $H_\text{S}$. The operators $G_m$ may include any combination of products of the Pauli matrices $\{\sigma_x, \sigma_y, \sigma_z\}$.}
	\label{fig:DAQC_Ex}
\end{figure}

The goal of DAQC protocols is to construct a propagator 
$U_\text{P} = e^{-i T H_\text{P}}$, where $H_\text{P}$ is a Hamiltonian containing one- and two-body terms, starting from another Hamiltonian $H_\text{S}$ provided by the experimental platform. The target propagator $U_\text{P}$ is synthesized through a sequence of conjugations of the source Hamiltonian $H_\text{S}$, which likewise contains one- and two-body interactions. In other words, $U_\text{P}$ is obtained as a product of exponentials of $H_\text{S}$ conjugated by local unitaries, as illustrated below:
\begin{equation}\label{eq:Up_frm_Hs}
	U_\text{P} = \prod_q G_q^\dagger e^{-it_q\, H_\text{S}} G_q =\prod_q  e^{-it_q \, G_q^\dagger H_\text{S} G_q },
\end{equation}
where the operators $G_q$ used to conjugate $H_\text{S}$ are local single-qubit unitaries. This construction results in a sequence in which $H_\text{S}$ evolves for intervals $t_q$, alternated with local pulses applied to one or several qubits, as illustrated in Fig.~\ref{fig:DAQC_Ex}. Such a sequence effectively reproduces the two-body interactions of $H_\text{P}$, while the single-body terms are implemented at the end of the protocol through single-qubit gates. For this protocol to be valid we require the source Hamiltonian to be compatible with the problem, that is, we require $H_\text{S}$ to have non-zero couplings between qubits that are also connected in $H_\text{P}$, ${h_\text{P}}^{ij}_{\mu\nu}\neq0\Rightarrow \exists {h_\text{S}}^{ij}_{\mu'\nu'}\neq0$.

These conjugations introduce a phase in each coupling, which in a two-level system corresponds to a rotation $e^{-i\pi}$. As an illustrative example, consider an Ising-type Hamiltonian $H_\text{S} = \sum h_\text{s}^{ij}\, \sigma_z^i \otimes \sigma_z^j$, and let the $k$th qubit be conjugated by $\sigma_x^k$. The resulting transformed Hamiltonian is  
\begin{equation}\label{ExH_S}
	H_\text{S}^{(k)} = \sigma_x^k\, H_\text{S}\, \sigma_x^k = \sum_{i<j} (-1)^{\delta_{ik}+\delta_{jk}}\, h_s^{ij}\sigma_z^i \otimes \sigma_z^j .
\end{equation}

We now aim to simulate the evolution generated by $H_\text{P} = \sum h_\text{p}^{ij}\, \sigma_z^i \otimes \sigma_z^j$ for a total time $T$, using $H_\text{S}$ as defined above. Since the Hamiltonians commute, Eq.~\eqref{eq:Up_frm_Hs} reduces to  
\begin{equation}\label{eq:SIS_Exp_Sum}
	\text{exp}\left(-iTH_\text{P}\right) =  \text{exp}\left(-i \sum_q t_q\, G_q^\dagger H_\text{S} G_q \right) .
\end{equation}

If the conjugated Hamiltonians did not commute, the same result could be obtained by employing a Trotter decomposition. This relation can be rewritten as a linear system of equations,  
\begin{equation}\label{eq:sist_M}
	M\, \vec{t} = T\, \overrightarrow{h_\text{p}/h_\text{s}},
\end{equation}
where $h_\text{p}$ and $h_\text{s}$ correspond with the vectors with the coupling constants corresponding with the source and problem Hamiltonians respectively. This system of equations is solved for the vector $\vec{t}$, yielding the time durations of each block $t_q$. The matrix $M$ encodes the sign changes introduced by the conjugations, with rows corresponding to couplings and columns to the effective sign of the couplings (generated by the local operators) at each step.

The protocol introduced in Ref.~\cite{ArbitraryDAQC} systematically constructs the matrix $M$ of Eq.~\eqref{eq:sist_M} from the complete family of conjugations of $H_\text{S}$ under the set of local single-qubit operators. This family comprises $2^{n-1}$ operators in the $ZZ$ case and $4^n$ operators for arbitrary Hamiltonians. The matrix $M$ can be decomposed up to $n+1$ blocks, ranging from $M_{[0]}$, which consists of a single column of ones corresponding to conjugation by the identity, to $M_{[n]}$, where all qubits are simultaneously conjugated by different combinations of Pauli operators. It has been shown in Ref.~\cite{ArbitraryDAQC} that this construction guarantees the existence of a positive-time solution.

\section{Generalizing the protocol to qudit Hamiltonians}
\label{sec:DAQCqudit}

We now extend the DAQC protocol of Ref.~\cite{ArbitraryDAQC} to arbitrary local dimension $d$, i.e., to qudit circuits. The generalization relies on the complete Weyl–Heisenberg operator basis, a generalized Clifford algebra defined as~\cite{Bertlmann2008}
\begin{equation}
	W_{nm}=Z^n X^m = \sum_{k=0}^{d-1} w^{kn} \,| k \rangle
	\langle (k+m) \,\textrm{mod}\,d|, 
\end{equation}
where $w = e^{2\pi i/d}$ and $n, m \in [0, d-1]$. Here, $W_{10}=Z$ and $W_{01}=X$ denote the phase- and position-shift operators, generalizing the Pauli matrices through the clock and shift formalism. The Weyl operators satisfy the following relations \cite{Popp2024}:
\begin{equation}
	\label{weylRelations}
	\begin{aligned}
		& W_{k_1,k_2}W_{l_1,l_2} = w^{l_1 k_2}~W_{k_1+l_1, k_2+l_2},  \\
		& W_{l_1,l_2}^\dagger = w^{l_1 l_2}~W_{-l_1, -l_2}.
	\end{aligned}
\end{equation}
From these properties, the conjugation of an element of the basis by another element of the basis yields
\begin{eqnarray}
	W_{k_1,k_2}^\dagger W_{l_1,l_2} W_{k_1,k_2}
	=
	w^{\,l_2k_1 - l_1k_2} \, W_{l_1,l_2},
	\label{eq:W_conjugation}
\end{eqnarray}
which reduces to a sign change for $d=2$, since $w = e^{i\pi} = -1$. 

In a similar manner to the qubit case, we can write any qudit two-body Hamiltonian in terms of this basis
\begin{equation}
H = \sum_{i,j} \sum_{l_1,l_2,l_3,l_4=0}^{d-1}
h_{\vec{\ell}}^{\,ij}\,
W^{(i)}_{l_1 l_2} \otimes W^{(j)}_{l_3 l_4},
\end{equation}
where $\vec{\ell}={l_1,l_2,l_3,l_4}$ specifies the operators in the coupling in the Weyl basis. As for the qubit case, we require the source and the problem Hamiltonians to be compatible. Focusing on a single coupling, the corresponding conjugation reads
\begin{multline}
	(W_{k_1,k_2}^\dagger \otimes W_{k_3,k_4}^\dagger)
	(W_{l_1,l_2} \otimes W_{l_3,l_4})
	(W_{k_1,k_2} \otimes W_{k_3,k_4})\\
	=
	(w^{l_2k_1 - l_1k_2} \, W_{l_1,l_2})\otimes	(w^{l_4k_3 - l_3k_4} \, W_{l_3,l_4}).
    \label{eq:WW_conjugation}
\end{multline}

The construction of the matrix $M$ proceeds analogously for any $d$: it is organized into blocks $M_{[k]}$, where $k$ denotes the number of qudits conjugated simultaneously. The matrix $M$ has up to $d^{2n}$ columns, while the number of rows depends on the couplings present in $H_\text{S}$ and $H_\text{P}$, which will be at most $\mathcal{O}(d^4n^2)$. As per Carathéodory's theorem, the number of analog blocks required is equal to the number of couplings in the system.  The extension of the proof to $d$-level systems is provided in Appendix~\ref{ap:ExistenceProof} and \ref{ap:ExistencePositive}, completing the protocol.

A particularly relevant case involves Hamiltonians expressed in the spin operator basis $\{S_x, S_y, S_z\}$, such that:
\begin{equation}
	H = \sum_{i, j, \mu, \nu} h_{\mu,\nu}^{i, j} \, S_\mu^{(i)} \otimes S_\nu^{(j)}.
\end{equation}
This basis exhibits $\mathrm{SU}(2)$ dynamics, requiring generators that introduce a phase $e^{i\pi}$, analogous to the qubit case. Conjugating an element $S_\mu$ of the basis by an operator generated by $S_\nu$ yields
\begin{multline}
	e^{i\theta S_\nu} S_\mu e^{-i\theta S_\nu}= S_\mu + \sum_{n=1}^{\infty} \frac{1}{n!} \underbrace{[S_\nu,\ldots[S_\nu, S_\mu]]\ldots]}_{\text{n times commutator}}\\ 
	=S_\mu \cos(\theta)+\epsilon_{\mu \nu \eta} S_\eta \sin(\theta),
	\label{eq:Spin_conjugation}
\end{multline}
where $\epsilon_{\mu \nu \eta}$ is the Levi-Civita symbol. Setting $\theta = \pi$ recovers the qubit equivalence. For such Hamiltonians, the $d=2$ matrix $M$ can be directly employed for any spin, provided that the operations $e^{-i\pi S_\nu}$ and their complex conjugates are available within the experimental platform. This reduces the complexity to $\mathcal{O}(n^2)$ and the number of columns to $4^n$, as with qubits.

\section{Example for qutrits}
\label{sec:DAQCqutritEx}

\begin{figure*}[t]
	\includegraphics[width=1\textwidth]{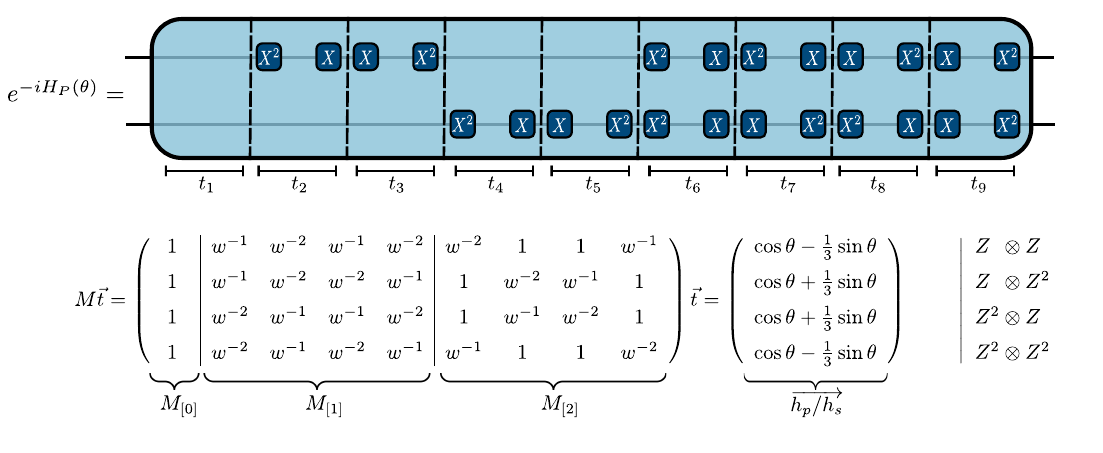}
	\caption{Analytical example of the simulation of $H_\text{P}(\theta)$ for $n = 2$. Each row $i$ of matrix $M$ identifies the two-body operator in the Hamiltonian, while each column $j$ specifies the phase acquired by the operator upon conjugation. The total number of digital–analog blocks shown here is illustrative; the final implementation will not require nine blocks.}
	\label{fig:DAQC_Ex_for_qutrits}
\end{figure*}

\begin{figure}[t]
	\includegraphics[width=0.5\textwidth]{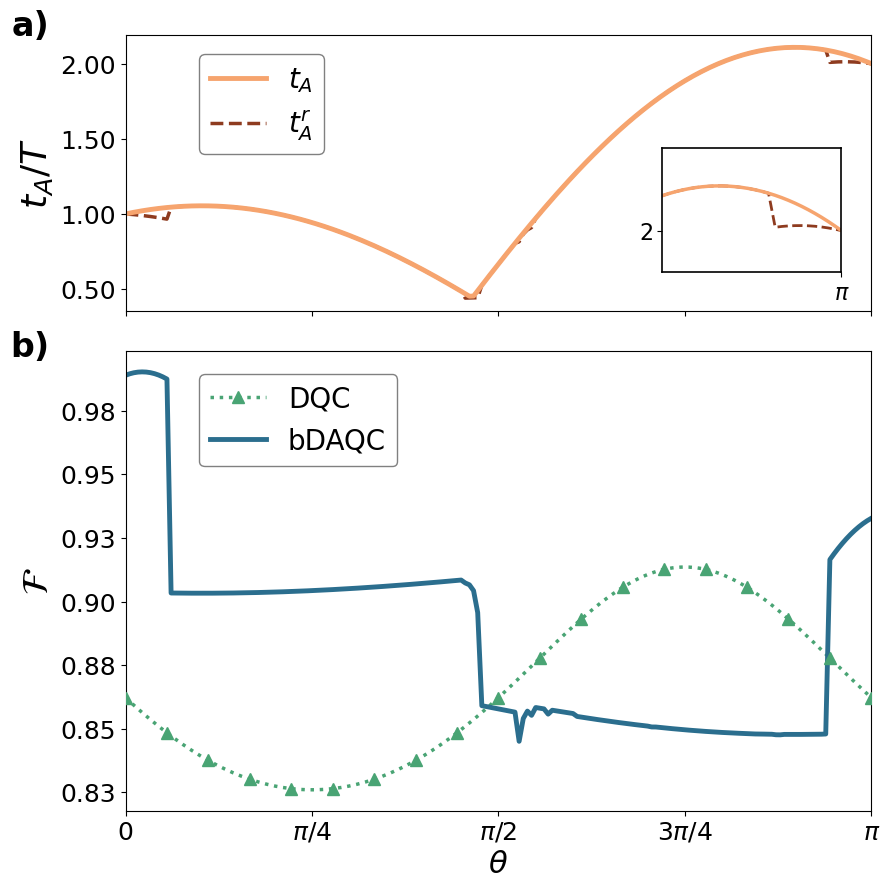}
	\caption{Total simulation time and fidelities for bDAQC and DQC. 
		Panel (a) compares the ideal total simulation time $t_A$ with the effective time $t^r_A$ after discarding blocks whose duration $t_i$ is too short to be implemented.
		Panel (b) shows the fidelity difference when simulating $H_\text{P}(\theta)$ using bDAQC and digital circuits. 
		Four regimes appear in bDAQC: the initial point $\theta = 0$ and its adjacent points, the region $\theta < \pi/2$, the interval $\pi/2 < \theta < \pi$, and the endpoint $\theta = \pi$ and its adjacent points. 
		The number of single-qutrit gates in these regimes is $0$, $18$, $27$, and $9$, respectively, which accounts for the observed variations in fidelity.}
	\label{fig:bDAQCvsDQC}
\end{figure}

To showcase the validity of the DAQC protocol, we consider a $d = 3$ qutrit example. Specifically, we simulate an isotropic bilinear–biquadratic (BLBQ) Ising-like Hamiltonian for spin $s = 1$
\begin{equation}
	\label{eq:BLBQ_HP}
	H_\text{P} = \sum_{i=1}^{n-1} \Big[ 
	\big( S_z^{(i)} S_z^{(i+1)} \big)\cos\theta
	+ \big( S_z^{(i)} S_z^{(i+1)} \big)^2 \sin\theta
	\Big],
\end{equation}
for a total time $T$, using the following source Hamiltonian:
\begin{equation}
	\label{eq:HS_form}
	H_\text{S} = \sum_{i=1}^{n-1} S_z^{(i)} S_z^{(i+1)}.
\end{equation}

Following the decomposition in Ref.~\cite{Bertlmann2008},
\begin{eqnarray}
	\ket{j}\bra{j} &=& \frac{1}{3}\sum_{l=0}^{2} w^{-lj}\, Z^l, 
	\qquad w = e^{\frac{2\pi i}{3}}, \nonumber\\
	&& Z = 
	\begin{pmatrix}
		1 & 0 & 0 \\
		0 & w & 0 \\
		0 & 0 & w^{-1}
	\end{pmatrix}.
\end{eqnarray}

The Hamiltonian operators in the Weyl–Heisenberg basis can be expressed as \footnote{From this point on, the indices $i$ and $i+1$ are omitted for simplicity.}
\begin{eqnarray}
	S_z = \dfrac{1}{3}  \Big[ (1-w^{-1}) Z + (1-w^{-2}) Z^2 \Big],
\end{eqnarray}
\begin{eqnarray}
	{S_z}^2 = \dfrac{1}{3}  \Big[ 2\, \mathbb{I} + (1+w^{-1}) Z + (1+w^{-2}) Z^2 \Big],
\end{eqnarray}
where $w = e^{i 2\pi /3}$. Since we are only interested in simulating the two-body terms, we redefine ${S_z}^2$ as $S_z^{'2} = {S_z}^2 - \frac{2}{3} \mathbb{I}$. The resulting two-body operators are therefore
\begin{multline}
	S_z  S_z = \frac{1}{9}\Big[
	(1-w^{-1})^2\, (Z \, Z)
	+ (1-w^{-2})^2\, (Z^2 \, Z^2)\\
	+ (1-w^{-1})(1-w^{-2})\, (Z \, Z^2+Z^2 \, Z) 
	\Big].
\end{multline}
\begin{multline}
	S_z^{'2} S_z^{'2} = \frac{1}{9}\Big[
	(1+w^{-1})^2 \,(Z \, Z)
	+ (1+w^{-2})^2 \,(Z^2 \, Z^2)\\
	+ (1+w^{-1})(1+w^{-2})\,(Z \, Z^2 + Z^2 \, Z)
	\Big].
\end{multline}
As an analytical example, we construct the matrix $M$ for $n = 2$, shown in Fig.~\ref{fig:DAQC_Ex_for_qutrits}.

\subsection{Numerical results}

For the simulations, we employ the banged-DAQC approach \cite{Parra2020}, in which single-qudit rotations are applied while the Hamiltonian remains active, rather than switching it on and off, as this is experimentally more feasible. The initial density matrix corresponds to a GHZ state, which is maximally entangled.

Regarding the parameters, the thermalization time is set to $T_1 = 100T$, while the duration of the single-qutrit gate $X$ is $\Delta t = 10^{-2}T$ with a fidelity of $\sim99.4\%$ \cite{Ringbauer2022}. When solving the linear system, solutions with $t_i < 4\Delta t$ are discarded, since these blocks would be shorter than the longest single-qutrit conjugation times.

In the digital case, two-qutrit controlled-phase gates are implemented using the unitary evolution from Ref.~\cite{Goss2022}, with a fixed fidelity of $95\%$, while single-qutrit gates are applied as detailed in Appendix~\ref{ap:DQC_setup}.

Figure~\ref{fig:bDAQCvsDQC} shows the fidelity difference between DAQC and digital simulations of $H_\text{P}(\theta)$ for $n = 6$ and angles $\theta \in [0, \pi]$. The bDAQC protocol achieves higher fidelities for $\theta < \pi/2$. For $\theta = 0$, no gates are needed, hence the almost perfect fidelity. At small $\theta$, blocks reproducing the exact Hamiltonian are too short to be implemented, so single-qutrit gates are omitted. The sharp decrease of fidelity afterwards is explained by the increase of single-qutrit gates, which are now a total of 18. In the region between $\pi/2$ and $\pi$, the decrease in fidelity results from an increased number of single-qutrit gates relative to the previous region, amounting to a total of 27. At points close to $\theta = \pi$, some blocks are too short and get discarded, dropping the number of single-qutrit gates to 9, improving the fidelity.

\section{Conclusions}
\label{sec:Conclusions}

In this work, we have extended the DAQC framework to qudits, enabling the simulation of any target Hamiltonian using a source Hamiltonian. This is achieved by conjugating the analog blocks with pairs of single-qudit gates selected from the Weyl–Heisenberg operator basis. The proposed method requires solving a complex linear system of $d^{2n}$  equations, but optimal solutions would only require the implementation of $\mathcal{O}(d^4 n^2)$ blocks. We have also demonstrated that the protocol originally formulated for qubits can be directly applied to higher-dimensional representations of $SU(2)$ spin systems. Furthermore, due to the mathematical structure of the phase matrix $M$, we expect that most of the properties and bounds derived for qubits can be straightforwardly extended to qudits.

We have illustrated the protocol by simulating the bilinear-biquadratic Ising-like model for spin $s=1$ using an Ising-like spin Hamiltonian as a resource Hamiltonian. We have considered both decoherence and gate errors, showing a comparable performance to a digital approach with a problem-tailored gate set. Additionally, the formalism enables the generation of new effective operators: distinct elements within a given basis that share identical Weyl components can be transformed into one another, as illustrated by the correspondence between spin and quadratic spin operators in our example simulating $\theta=\pi/2$.

The theory developed here expands the range of systems that can be efficiently simulated with digital-analog protocols, for instance those containing quadrupolar terms. Additionally, qudit DAQC would eliminate the need for multi-qubit mappings in bosonic simulations~\cite{PhysRevLett.129.160501, PhysRevResearch.6.013202} while showing higher resilience against noises. Consequently, this paves the way for simulating gauge theories requiring fewer quantum resources, an essential advantage in the NISQ era. 

The generalization to qudits provides a comprehensive experimental framework to implement DAQC circuits employing an entangling Hamiltonian and only the single-qudit operators $\{X_d, Z_d\}$. This makes this approach implementable in state-of-the-art platforms, such as $d$-level system ion-traps \cite{Ringbauer2022}.

Future efforts towards broadening the DAQC approach should focus on mitigating Trotter errors, which now scale not only  with the number of qudits but also with their dimensionality \footnote{The total Trotter error is upper-bounded by $\lVert H_{\text{S}} \rVert$}. Developing systematic strategies to minimize these errors will be crucial for realizing increasingly scaling DAQC algorithms.

\begin{acknowledgments}
The authors thank David Aguirre and Yongcheng Ding for insightful and helpful discussions
and acknowledge financial support from OpenSuperQ+100 (Grant No. 101113946) of the EU Flagship on Quantum Technologies,
from Project Grant No. PID2024-156808NB-I00,
also from Spanish Ramón y Cajal Grant No. RYC-2020-030503-I funded by MICIU/AEI/10.13039/501100011033 and by “ERDF A way of making Europe” and “ERDF Invest in your Future”, and from the Spanish Ministry of Science and Innovation under the Recovery, Transformation and Resilience Plan (CUCO, MIG-20211005).
This project has also received support from the Spanish Ministry for Digital Transformation and of Civil Service of the Spanish Government through the QUANTUM ENIA project call - Quantum Spain, and by the EU through the Recovery, Transformation and Resilience Plan – NextGenerationEU within the framework of the Digital Spain 2026 Agenda.
We acknowledge funding from Basque Government through Grant No. IT1470-22,
and the IKUR Strategy under the collaboration agreement between Ikerbasque Foundation and BCAM on behalf of the Department of Education of the Basque Government. 
AA acknowledges grant PID2021-125823NA-I00 funded by MICIU/AEI/10.13039/501100011033 and by the “FSE+".

\end{acknowledgments}

\appendix

\section{Proof of existence of a solution}
\label{ap:ExistenceProof}

To demonstrate the existence of a solution for the matrix $M$, we focus on the $M_{[2]}$ block, corresponding to the columns representing all possible conjugations of the Hamiltonian using two single-qudit gates simultaneously. The structure of the submatrices composing this block is given by

\begin{equation}
    \label{submatrices}
    \begin{aligned}
          & M^2 = (w^{l_2k_1-l_1k_2+l_4k_3-l_3k_4})_{ij} \\
          & M^{1.1} = (w^{l_2k_1-l_1k_2})_{ij} \\
          & M^{1.2} = (w^{l_4k_3-l_3k_4})_{ij} \\
          & M^0 = 1_{ij},
    \end{aligned}
\end{equation}
where $M^2$, $M^{1.1}$, $M^{1.2}$, and $M^0$ are matrices of dimension $(d^2-1)^2$, representing the addition of a phase to both qudits in the coupling, to the first qudit, to the second qudit, and to none, respectively. The indices $l_m$ label the rows and the indices $k_m$ the columns, thus

\begin{equation}
	\left.\begin{aligned}
		d \cdot l_1 + l_2 &= \alpha \\
		d \cdot l_3 + l_4 &= \beta
	\end{aligned}
	\right\}
	\quad
	\begin{aligned}
		(d^2 - 1)(\alpha - 1) + \beta = i.
	\end{aligned}
\end{equation}

\begin{equation}
	\left.\begin{aligned}
		d \cdot k_1 + k_2 &= \eta \\
		d \cdot k_3 + k_4 &= \nu
	\end{aligned}
	\right\}
	\quad
	\begin{aligned}
		(d^2 - 1)(\eta - 1) + \nu = j
	\end{aligned}.
\end{equation}

The indices $\{l_1, l_2, l_3, l_4\}$ are ordered such that the first element is $\{0, 1, 0, 1\}$, since $l_1 = 0$ implies $l_2 \neq 0$, and $l_3 = 0$ implies $l_4 \neq 0$ (as no operator in a two-body term can be the identity).  All indices $l_m$ take values from $0$ to $d-1$, starting from $l_4$ and ending at $l_1$.  Their relation to the row index $i$ is given by
\begin{eqnarray}
	l_1 &=& 
	\left\lfloor 
	\frac{\,\left\lfloor (i-1)/(d^{2}-1) \right\rfloor + 1\,}{d}
	\right\rfloor
	\\[6pt]
	l_2 &=& 
	\left\lfloor \frac{i-1}{d^{2}-1} \right\rfloor + 1
	\nonumber\\
	&&\quad
	- d\left\lfloor 
	\frac{\,\left\lfloor (i-1)/(d^{2}-1) \right\rfloor + 1\,}{d}
	\right\rfloor
	\\[6pt]
	l_3 &=& 
	\left\lfloor 
	\frac{
		i -
		\left( \left\lfloor (i-1)/(d^{2}-1) \right\rfloor \right)(d^{2}-1)
	}{d}
	\right\rfloor
	\\[6pt]
	l_4 &=& 
	i -
	\left\lfloor \frac{i-1}{d^{2}-1} \right\rfloor (d^{2}-1)
	\nonumber\\
	&&\quad
	- d\left\lfloor
	\frac{
		i -
		\left( \left\lfloor (i-1)/(d^{2}-1) \right\rfloor \right)(d^{2}-1)
	}{d}
	\right\rfloor,
\end{eqnarray}
and the same convention applies to $k_m$, where the correspondence with the column index $j$ follows identically after replacing $i \rightarrow j$.

  The distribution of these submatrices remains the same for all $d$, changing only in size and internal structure. The properties of $\{M^2, M^{1.1}, M^{1.2}, M^0\}$ are summarized as follows:

\begin{subequations}
\label{subMpp}
\begin{align}
&\{M^2, M^{1.1}, M^{1.2}, M^0\} \text{ commute in pairs.} \label{subMpp1} \\
&M^{1.1}M^{1.2} = M^2 M^0 = M^0 \label{subMpp2} \\
&M^{1.1}M^0 = M^{1.2}M^0 = -(d^2 - 1)M^0 \label{subMpp3} \\
&(M^0)^2 = (d^2 - 1)^2 M^0 \label{subMpp4} \\
&(M^{1.1})^k = (-1)^{k - 1}(d^2 - 1)^{k - 1}(M^2)^{k - 1}M^{1.1} \label{subMpp5} \\
&(M^{1.2})^k = (-1)^{k - 1}(d^2 - 1)^{k - 1}(M^2)^{k - 1}M^{1.2}, \label{subMpp6}
\end{align}
\end{subequations}
and their eigenvalues, denoted as $\{\lambda_2, \lambda_{1.1}, \lambda_{1.2}, \lambda_0\}$, are given by $\lambda_2 \in \{1, \pm d, \pm d^2\}$, $(\lambda_{1.1}, \lambda_{1.2}) \in \{0, \pm (d^2-1), \pm d(d^2-1)\}$, and $\lambda_0 \in \{0, (d^2-1)^2\}$. Proofs of these properties are provided in the Supplemental Material.

\subsection{$M_{[2]}$ uniqueness}

We now prove that the determinant of $M_{[2]}$ is nonzero, ensuring the existence of a solution. To this end, we turn to M. H. Ingraham’s note on determinants~\cite{APX:Ingraham1937}, which states that if a square matrix $A = (A_{ij})$ is composed of square submatrices $A_{ij}$ that commute pairwise, then the determinant of the matrix constructed by taking $\det(A_{ij})$ as its elements is equal to $\det(A)$. For illustration, consider the $2\times2$ case
\begin{equation} 
	\text{det}(A) = 
	\begin{vmatrix} 
		A_{11} & A_{12} \\
		A_{21} & A_{22} 
	\end{vmatrix} 
	= A_{11}A_{22} - A_{12}A_{21},
\end{equation}
where the identity holds when all $A_{ij}$ commute in pairs.

Hence, we evaluate $\det M_{[2]}$ through its mutually commuting submatrices (see Eq.~\ref{subMpp1}). Using Eqs.~\ref{subMpp2} and \ref{subMpp3}, the cross terms can be grouped into the set of matrices $\{M^2, M^0, M^2M^{1.1}, M^2M^{1.2}\}$, as all products are powers of these matrices, as in Ref.~\cite{ArbitraryDAQC}. Consequently, the determinant takes the form of Eq.~(\ref{eq:DetM2_sim}),
\begin{widetext}
\begin{eqnarray}
	\text{det}(M_{[2]})
	&=&
	\Big|
	(M^2)^N
	+ \alpha M^0
	+ \displaystyle\sum_{k=2}^{N-1}
	\Big[
	\beta_k (M^2)^{N-k}(M^{1.1})^k
	+ \gamma_k (M^2)^{N-k}(M^{1.2})^k
	\Big]
	\Big|
	\nonumber\\
	&=&
	\Big|
	(M^2)^N
	+ \alpha M^0
	+ (M^2)^{N-1}
	\displaystyle\sum_{k=2}^{N-1}
	(-1)^{k-1} \Big[ \beta_k(d^2 - 1)^{k-1} (M^{1.1}) 
	+ 
	\gamma_k(d^2 - 1)^{k-1} (M^{1.2}) \Big]
	\Big|
	\label{eq:DetM2_1sim}
	\nonumber\\
	&=&
	\Big|
	(M^2)^{N-1}
	\Big|
	\;
	\Big|
	M^2
	+ \alpha M^0
	+ 
	\displaystyle\sum_{k=2}^{N-1}
	(-1)^{k-1} \Big[ \beta_k(d^2 - 1)^{k-1} (M^{1.1}) 
	+ 
	\gamma_k(d^2 - 1)^{k-1} (M^{1.2}) \Big]
	\Big|
	\label{eq:DetM2_2sim}
	\nonumber\\
	&=&
	\displaystyle\prod_i
	\Big|
	(\lambda_i^2)^{N-1}
	\Big|
	\cdot
	\Big|
	\lambda_i^2
	+ \alpha \lambda_i^0
	+ \displaystyle\sum_{k=2}^{N-1}
	(-1)^{k-1} \Big[ \beta_k(d^2 - 1)^{k-1} (\lambda_i^{1.1}) 
	+ 
	\gamma_k(d^2 - 1)^{k-1} (\lambda_i^{1.2}) \Big]
	\Big|,
	\label{eq:DetM2_sim}
\end{eqnarray}
\end{widetext}
where $\alpha, \beta_k, \gamma_k\in\mathbb{Z}$. The first equality is obtained by applying properties (\ref{subMpp5}) and (\ref{subMpp6}), whereas the second one follows from Eq.~(\ref{subMpp2}) after factoring out common terms. Finally, the last equality arises from the commutation of all matrices, where the determinant of the sum equals the sum of the eigenvalues.

By applying a modular argument to Eq.~(\ref{eq:DetM2_sim}), we can readily show that the determinant is nonzero:
\begin{multline}
    \lbrack \text{Det}(M_{[2]})\rbrack \text{mod}(d^2-1)
    =\prod_i \lbrack|(\lambda_i^2)^{N}|\rbrack  \text{mod}(d^2-1) \neq 0
\end{multline}

\section{Proof of existence of a positive time solution}
\label{ap:ExistencePositive}

Proving the uniqueness of the block $M_{[2]}$ allows us to define an $N$-polytope, since there are $N$ linearly independent vectors in $M$. We now show that the barycenter of this polytope lies at $\vec{0} = (0, \ldots, 0)$ such that
\begin{equation}
    \sum_{q=1}^{d^{2n}}M_q= \begin{pmatrix}
        0 \\
        \vdots\\
        0
    \end{pmatrix}.
\end{equation}
To this end, we review all possible configurations in the summation for a given row, through the following analogy:
\begin{itemize}
	\item On one hand, we consider all possible conjugations for a qudit involved in the coupling, which are represented by the following sum, where $l_i$ are fixed:
\begin{equation}
    \sum_{k_1=0}^{d-1}\sum_{k_2=0}^{d-1} w^{l_2k_1-l_1k_2} - 1 =-1.
\end{equation}
This can be regarded as flipping the switch on qudit $i$ or $j$ within the coupling $ij$, which contributes $-1$ to the sum.

\item On the other hand, flipping the switch on a qudit that is not part of the coupling contributes $+1$ a total of $d^2 - 1$ times, accounting for all possible conjugations with the basis elements.
\end{itemize}

If we flip the switch on $k$ qudits, among which $m$ belong to the pair $ij$, the total contribution to the sum becomes:
\begin{equation}
    \binom{2}{m}(-1)^m \times \binom{n-2}{k - m}(d^2 - 1)^{k - m}
\end{equation}
where $\binom{2}{m}$ represents all possible ways to choose $m$ switches within the pair, and $\binom{n - 2}{k - m}$ represents the ways to choose the remaining $n - 2$ switches. The sum over all columns for each row is thus given by:
\begin{multline}
    S =  \sum_{m=0}^{2} \binom{2}{m} (-1)^m 
    \times \sum_{k=m}^{n-2+m}\binom{n-2}{k-m}  (d^2 - 1)^{k-m}.
\end{multline}

Let us evaluate this sum over all possible $m$ values:
\begin{itemize}
    \item $m=0$ case:
    \begin{eqnarray}
         \sum_{k=0}^{n-2}  \binom{2}{0} \binom{n-2}{k}  (d^2 - 1)^{k}=(d^2)^{n-2}
    \end{eqnarray}
    \item $m=1$ case:
    \begin{eqnarray}
         \sum_{k=1}^{n-1}  \binom{2}{1} \binom{n-2}{k-1}  (-1)(d^2 - 1)^{k-1}=-2 (d^2)^{n-2}
    \end{eqnarray}
    \item $m=2$ case:
    \begin{eqnarray}
         \sum_{k=2}^{n}  \binom{2}{2} \binom{n-2}{k-2}  (d^2 - 1)^{k-2}=(d^2)^{n-2}
    \end{eqnarray}
\end{itemize}

Hence, the total sum for each row can be expressed as:
\begin{eqnarray}
    S = d^{2(n-2)} - 2d^{2(n-2)} + d^{2(n-2)} = 0.
\end{eqnarray}
Following this reasoning, we can analogously construct a polytope as in Ref.~\cite{ArbitraryDAQC} with the matrix columns as the vertices. Then, we can find a hypersphere centered at the origin and which is completely inside the polytope, allowing us to prove the existence of a positive solution to the problem.

\section{DQC circuit to simulate $H_\text{P}(\theta)$} 
\label{ap:DQC_setup}

\begin{figure}[b]
	\includegraphics[width=0.31\textwidth]{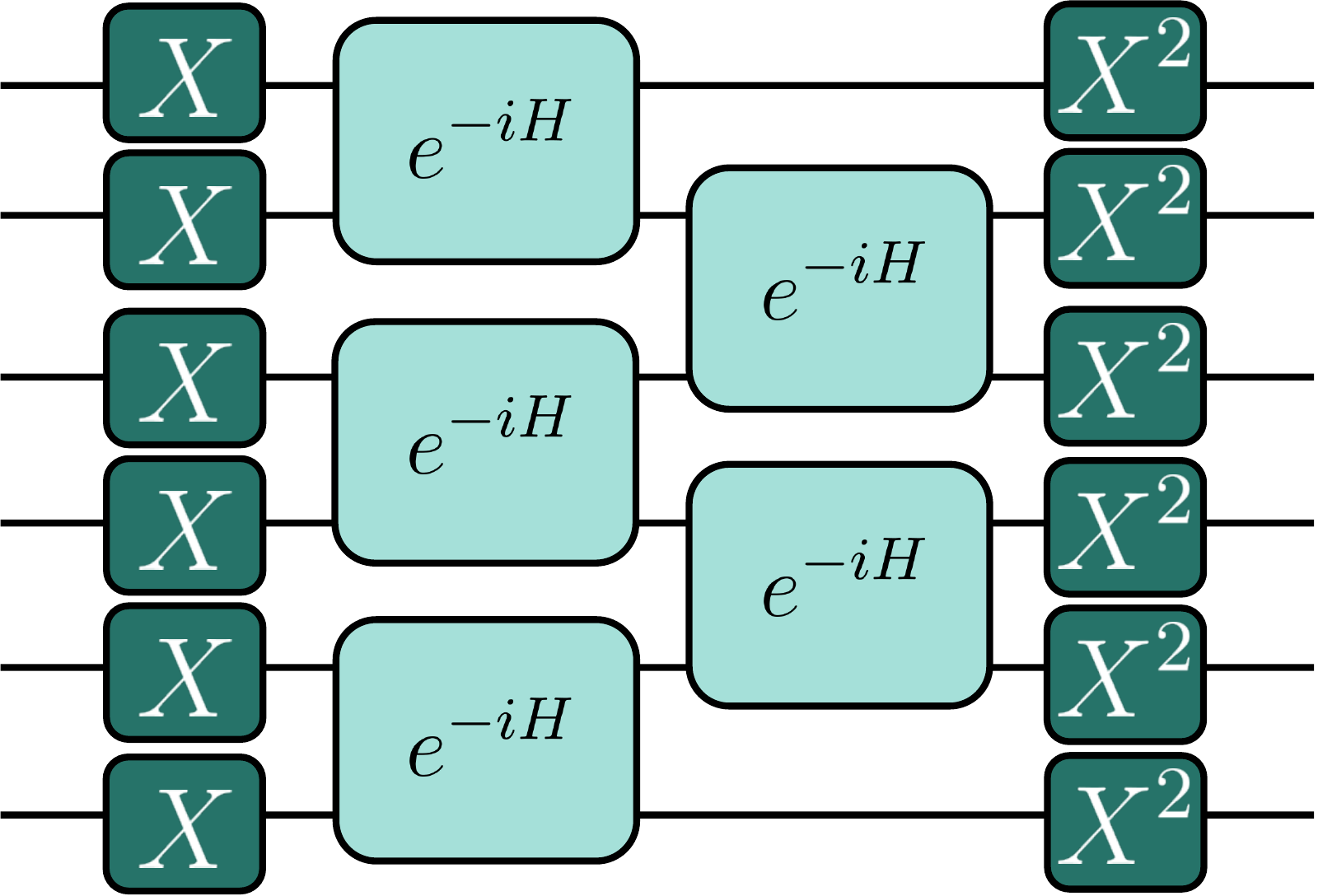}
	\caption{Brick-wall circuit structure used to simulate $H_\text{P}(\theta)$ with the Hamiltonian $H$ for $n = 6$.}
	\label{fig:DQC_circuitZZ}
\end{figure}

We consider the Hamiltonian generating the unitary operator that can be implemented as:
\begin{multline}
	H = (\alpha_{11}+\alpha_{22})\,\tau
	\big(\ketbra{11}{11}+\ketbra{22}{22}\big) \\
	 \quad + (\alpha_{12}+\alpha_{21})\,\tau
	\big(\ketbra{12}{21}+\ketbra{21}{12}\big),
\end{multline}
where this Hamiltonian and the parameters $\alpha$ can be found in Ref.~\cite{Goss2022}. If we conjugate it, we obtain the isotropic bilinear–biquadratic (BLBQ) Ising-like Hamiltonian:
\begin{multline}
	(X \otimes X)\, H \,(X^2 \otimes X^2) 
	= \big(S_z S_z\big)\cos\theta	 + \big(S_z S_z\big)^2\sin\theta,
\end{multline}
where 
\begin{eqnarray}
	\cos\theta &=& \left( \frac{\alpha_{11} + \alpha_{22} + \alpha_{12} + \alpha_{21}}{2} \right)\,\tau \\
	\sin\theta &=& \left( \frac{\alpha_{11} + \alpha_{22} - \alpha_{12} - \alpha_{21}}{2} \right)\,\tau.
\end{eqnarray}

For the simulation, we assume that the parameters can be approximated in such a way that both values are achieved, and we fix the fidelity of this two-qutrit gate at $95\%$, inspired by the fidelity of the $CZ$ gate. To minimize error propagation, the circuit is constructed in a brick-wall configuration~\cite{Tepaske2024}, as shown in Fig.~\ref{fig:DQC_circuitZZ}.

\bibliography{apssamp}

\include{SM}

\end{document}

%% file: SM.tex
\onecolumngrid

\section*{Supplementary Material}

The objective of DAQC protocols is to engineer a propagator $U_{\mathrm{P}} = e^{-i T H_\text{P}}$, where $H_\text{P}$ is a Hamiltonian composed of one- and two-body terms, starting from a source Hamiltonian $H_\text{S}$ supplied by the experimental platform. The target propagator $U_{\mathrm{P}}$ is synthesized through a sequence of conjugations of the source Hamiltonian $H_\text{S}$, which likewise contains one- and two-body interactions. In essence, $U_{\mathrm{P}}$ is obtained as a product of exponentials of $H_\text{S}$ conjugated by local unitaries, as illustrated below:
\begin{equation}\label{Apeq:Up_frm_Hs}
    U_{\mathrm{P}} = \prod_q G_q^\dagger e^{-i t_q\, H_\text{S}} G_q
    = \prod_q e^{-i t_q \, G_q^\dagger H_\text{S} G_q } ,
\end{equation}
where the operators $G_q$ that conjugate $H_\text{S}$ are local single-qudit unitaries. This construction yields a sequence in which the system evolves under $H_\text{S}$ for intervals $t_q$, alternated with local pulses applied to one or several qudits.

Both the target Hamiltonian $H_\text{P}$ and the source Hamiltonian $H_\text{S}$ can be expressed in the Weyl-Heisenberg operator basis. Their general form is
\begin{equation}
H = \sum_{i,j} \sum_{l_1,l_2,l_3,l_4=0}^{d-1}
h_{\vec{\ell}}^{\,ij}\,
W^{(i)}_{l_1 l_2} \otimes W^{(j)}_{l_3 l_4},
\end{equation}
where $W^{(i)}_{l_1 l_2}$ and $W^{(j)}_{l_3 l_4}$ denote Weyl-Heisenberg operators acting on qudits $i$ and $j$ respectively, and $\vec{\ell}={l_1,l_2,l_3,l_4}$ specifies the operators.  
The coupling coefficients $h_{\vec{\ell}}^{\,ij}$ are, in general, complex for dimension $d>2$. The conjugating operators $G_q$ consist of Weyl-Heisenberg single-body operations applied to $k$ qudits, e.g., if only one qudit is being conjugated, $G_q = W^{(q)}_{k_1 k_2}$. These conjugations modify the phases of the individual couplings. By equating Eq.~\eqref{Apeq:Up_frm_Hs} with the exponential form of the target evolution, we obtain
\begin{equation}\label{Apeq:SIS_Exp_Sum}
    \exp\!\left(-i T H_\text{P}\right)
    = 
    \exp\!\left(-i \sum_q t_q\, G_q^\dagger H_\text{S} G_q \right).
\end{equation}

If the conjugated Hamiltonians do not commute, the same construction can be achieved by approximating it via Trotter decomposition. The above identity can be written as a linear system,
\begin{equation}
    M\, \vec{t}
    = 
    T\, \overrightarrow{h_\text{p} / h_\text{s}},
\end{equation}
which is then solved for the vector $\vec{t}$ that specifies the duration of each block $t_q$. The matrix $M$ captures the phase factors added by the conjugations: its rows correspond to the various couplings, while its columns encode the phase induced by the local unitary operations at each step.

We establish the existence of a solution for $M$ by showing that the determinant of the submatrix $M_{[2]}$ is nonzero. Here $M_{[2]}$ denotes the block of $M$ corresponding to all configurations of conjugations of the Hamiltonian with two single-qudit gates. The submatrices composing this block have the structure
\begin{equation}
\label{submatrices}
\begin{aligned}
    & M^{2} = \big(w^{\,l_2k_1 - l_1k_2 + l_4k_3 - l_3k_4}\big)_{ij} \\
    & M^{1.1} = \big(w^{\,l_2k_1 - l_1k_2}\big)_{ij} \\
    & M^{1.2} = \big(w^{\,l_4k_3 - l_3k_4}\big)_{ij} \\
    & M^{0} = 1_{ij}
\end{aligned}
\end{equation}
where $M^{2}$, $M^{1.1}$, $M^{1.2}$, and $M^{0}$ are $(d^{2}-1)^2$-dimensional matrices representing, respectively, the application of a phase to both qudits in the interaction, to the first qudit only, to the second qudit only, or to neither. The indices $l_m$ and $k_m$ label the rows and columns respectively, such that
\begin{equation}
	\left.\begin{aligned}
		d \cdot l_1 + l_2 &= \alpha \\
		d \cdot l_3 + l_4 &= \beta
	\end{aligned}
	\right\}
	\quad
	\begin{aligned}
		(d^2 - 1)(\alpha - 1) + \beta = i,
	\end{aligned}
\end{equation}
\begin{equation}
	\left.\begin{aligned}
		d \cdot k_1 + k_2 &= \eta \\
		d \cdot k_3 + k_4 &= \nu
	\end{aligned}
	\right\}
	\quad
	\begin{aligned}
		(d^2 - 1)(\eta - 1) + \nu = j.
	\end{aligned}
\end{equation}

The indices $\{l_1, l_2, l_3, l_4\}$ are ordered so that the leading pattern is $\{0,1,0,1\}$. This reflects the constraints that $l_1 = 0$ requires $l_2 \neq 0$ and $l_3 = 0$ requires $l_4 \neq 0$, since no operator in a two-body term may be the identity. Each $l_m$ ranges from $0$ to $d-1$, with the ordering running from $l_4$ to $l_1$. Their mapping to the row index $i$ is
\begin{align}
	l_1 &= 
	\left\lfloor 
	\frac{\,\left\lfloor (i-1)/(d^{2}-1) \right\rfloor + 1\,}{d}
	\right\rfloor, \\[6pt]
	l_2 &= 
	\left\lfloor \frac{i-1}{d^{2}-1} \right\rfloor 
	+ 1
	- d\left\lfloor 
	\frac{\,\left\lfloor (i-1)/(d^{2}-1) \right\rfloor + 1\,}{d}
	\right\rfloor, \\[6pt]
	l_3 &= 
	\left\lfloor 
	\frac{
		i - 
		\left( \left\lfloor (i-1)/(d^{2}-1) \right\rfloor \right)(d^{2}-1)
	}{d}
	\right\rfloor, \\[6pt]
	l_4 &= 
	i - 
	 \left\lfloor \frac{i-1}{d^{2}-1} \right\rfloor (d^{2}-1)
	- d\left\lfloor 
	\frac{
		i - 
		\left( \left\lfloor (i-1)/(d^{2}-1) \right\rfloor \right)(d^{2}-1)
	}{d}
	\right\rfloor.
\end{align}
For the $k_m$ indices, it is sufficient to replace the row index $i$ with the column index $j$.

We review some of the properties regarding $M_{[2]}$ and $M^{2}$, $M^{1.1}$, $M^{1.2}$,  $M^{0}$ matrices in the following sections.

\subsection{Properties of the submatrices}
Some properties of the sub-block matrices are as follows:
\begin{enumerate}[label=(S.\arabic*)]
	\item \label{S1}  $\{M^2,M^{1.1},M^{1.2},M^0\}$ commute in pairs,
	\item \label{S2} $M^{1.1}M^{1.2}=M^2M^0=M^0$,
	\item \label{S3} $M^{1.1}M^0=M^{1.2}M^0=-(d^2-1)M^0$,
	\item \label{S4} $(M^0)^2=(d^2-1)^2M^0$,
	\item \label{S5} $(M^{1.1})^k=(-1)^{k-1}(d^2-1)^{k-1}(M^2)^{k-1}M^{1.1}$,
	\item \label{S6} $(M^{1.2})^k=(-1)^{k-1}(d^2-1)^{k-1}(M^2)^{k-1}M^{1.2}$.
\end{enumerate}

To prove these properties, we first analyze the following geometric series:
\begin{equation}
	\displaystyle \sum_{k=0}^{d-1}w^{km}=\dfrac{1-w^{md}}{1-w^m}=0 \quad \text{if\;}\; m \neq 0,
\end{equation}
which recursively gives:
\begin{equation} \label{fund}
	\displaystyle \sum_{k_1=0}^{d-1}\sum_{k_2=0}^{d-1}w^{l_2k_1-l_1k_2}=\sum_{k_1=0}^{d-1}w^{l_2k_1}\sum_{k_2=0}^{d-1}w^{-l_1k_2}=0,
\end{equation}
since the index pairs $(l_1, l_2)$, $(l_3, l_4)$, $(k_1, k_2)$, and $(k_3, k_4)$ can only simultaneously take the value $0$ if they correspond to the identity operator.

\subsubsection{Proof of (S.1)}

In the proofs of these properties, whenever we consider the product of two matrices, the indices of the first matrix will be given by the set $l_m$ for the rows and the set $k_m$ for the columns. The indices of the second matrix will be given by $p_m$ for the rows and $q_m$ for the columns. As an illustrative example of Eq.~(\ref{fund}), let us compute a representative element of the product $M^2 M^0$.
\begin{equation}\label{M2M0example}
	M^2 M^0 = A 
	\;\rightarrow\;
	A_{i\, j}
	=
	\Bigg(
	\displaystyle\sum_{k_1=0}^{d-1}
	\displaystyle\sum_{k_2=0}^{d-1}
	w^{l_2k_1 - l_1k_2}
	- 1
	\Bigg)
	\Bigg(
	\displaystyle\sum_{k_3=0}^{d-1}
	\displaystyle\sum_{k_4=0}^{d-1}
	w^{l_4k_3 - l_3k_4}
	- 1
	\Bigg).
\end{equation}
Here, the $l_i$ are fixed and depend on $i$ $\rightarrow 
\displaystyle S=\sum_{k_1=0}^{d-1}\sum_{k_2=0}^{d-1}w^{l_2k_1-l_1k_2}=0, $ due to $\displaystyle \sum_{k=0}^{d-1}w^{km}=\dfrac{1-w^{md}}{1-w^m}=0$ if $m \neq0$ and it is not possible for the pairs $(l_1,  l_2)$ and $(l_3, l_4)$ (nor any entire quartet of indices) to be simultaneously zero. Therefore:
\begin{equation*}
	A_{ij}=(-1)(-1)=1, \quad \forall i,j.
\end{equation*}

\noindent\rule{\linewidth}{0.4pt}

\begin{eqnarray}
	(M^2M^0)_{ij} &=&
	\Bigg( 
	\displaystyle\sum_{k_1=0}^{d-1} 
	\displaystyle\sum_{k_2=0}^{d-1} 
	w^{l_2k_1 - l_1k_2} - 1 
	\Bigg)
	\Bigg(
	\displaystyle\sum_{k_3=0}^{d-1} 
	\displaystyle\sum_{k_4=0}^{d-1} 
	w^{l_4k_3 - l_3k_4} - 1 
	\Bigg)
	= (-1)(-1) = 1, \quad \forall i,j,
	\label{eq:M2M0}
	\\[1.5ex]
	(M^0M^2)_{ij} &=&
	\Bigg( 
	\displaystyle\sum_{p_1=0}^{d-1} 
	\displaystyle\sum_{p_2=0}^{d-1} 
	w^{p_2q_1 - p_1q_2} - 1 
	\Bigg)
	\Bigg(
	\displaystyle\sum_{p_3=0}^{d-1} 
	\displaystyle\sum_{p_4=0}^{d-1} 
	w^{p_4q_3 - p_3q_4} - 1 
	\Bigg)
	= (-1)(-1) = 1, \quad \forall i,j.
	\label{eq:M0M2}
\end{eqnarray}

\noindent\rule{\linewidth}{0.4pt}

\begin{eqnarray}
	(M^{1.1}M^0)_{ij} &=&
	\Bigg( 
	\displaystyle\sum_{k_1=0}^{d-1}
	\displaystyle\sum_{k_2=0}^{d-1}
	w^{l_2k_1 - l_1k_2} - 1 
	\Bigg)
	\Bigg(
	\displaystyle\sum_{k_3=0}^{d-1}
	\displaystyle\sum_{k_4=0}^{d-1}
	1 - 1 
	\Bigg)
	= (-1)(d^2 - 1), \quad \forall i,j,
	\label{eq:M11M0}
	\\[1.5ex]
	(M^0M^{1.1})_{ij} &=&
	\Bigg( 
	\displaystyle\sum_{p_1=0}^{d-1}
	\displaystyle\sum_{p_2=0}^{d-1}
	w^{p_2q_1 - p_1q_2} - 1 
	\Bigg)
	\Bigg(
	\displaystyle\sum_{p_3=0}^{d-1}
	\displaystyle\sum_{p_4=0}^{d-1}
	1 - 1 
	\Bigg)
	= (-1)(d^2 - 1), \quad \forall i,j.
	\label{eq:M0M11}
\end{eqnarray}

\noindent\rule{\linewidth}{0.4pt}

\begin{eqnarray}
	(M^{1.2}M^0)_{ij} &=&
	\Bigg(
	\displaystyle\sum_{k_1=0}^{d-1}
	\displaystyle\sum_{k_2=0}^{d-1}
	1 - 1
	\Bigg)
	\Bigg(
	\displaystyle\sum_{k_3=0}^{d-1}
	\displaystyle\sum_{k_4=0}^{d-1}
	w^{l_4k_3 - l_3k_4} - 1
	\Bigg)
	= (d^2 - 1)(-1), \quad \forall i,j,
	\label{eq:M12M0}
	\\[1.5ex]
	(M^0M^{1.2})_{ij} &=&
	\Bigg(
	\displaystyle\sum_{p_1=0}^{d-1}
	\displaystyle\sum_{p_2=0}^{d-1}
	1 - 1
	\Bigg)
	\Bigg(
	\displaystyle\sum_{p_3=0}^{d-1}
	\displaystyle\sum_{p_4=0}^{d-1}
	w^{p_4q_3 - p_3q_4} - 1
	\Bigg)
	= (d^2 - 1)(-1), \quad \forall i,j.
	\label{eq:M0M12}
\end{eqnarray}

\noindent\rule{\linewidth}{0.4pt}

\begin{eqnarray}
	(M^2M^{1.1})_{ij} &=&
	\Bigg(
	\displaystyle\sum_{k_1=0}^{d-1}
	\displaystyle\sum_{k_2=0}^{d-1}
	w^{l_2k_1 - l_1k_2}
	\;w^{p_2q_1 - p_1q_2}
	- 1
	\Bigg)
	\Bigg(
	\displaystyle\sum_{k_3=0}^{d-1}
	\displaystyle\sum_{k_4=0}^{d-1}
	w^{l_4k_3 - l_3k_4}
	- 1
	\Bigg)
	\nonumber\\
	&=&
	\Bigg(
	\displaystyle\sum_{k_1=0}^{d-1}
	\displaystyle\sum_{k_2=0}^{d-1}
	w^{l_2k_1 - l_1k_2}
	\;w^{k_2q_1 - k_1q_2}
	- 1
	\Bigg)
	\Bigg(
	\displaystyle\sum_{k_3=0}^{d-1}
	\displaystyle\sum_{k_4=0}^{d-1}
	w^{l_4k_3 - l_3k_4}
	- 1
	\Bigg)
	\nonumber\\
	&=&
	\Bigg(
	\displaystyle\sum_{k_1=0}^{d-1}
	\displaystyle\sum_{k_2=0}^{d-1}
	w^{l_2k_1 - l_1k_2}
	\;w^{k_2q_1 - k_1q_2}
	- 1
	\Bigg)(-1),
	\label{eq:M2M11}
	\\[2ex]
	(M^{1.1}M^2)_{ij} &=&
	\Bigg(
	\displaystyle\sum_{k_1=0}^{d-1}
	\displaystyle\sum_{k_2=0}^{d-1}
	w^{l_2k_1 - l_1k_2}
	\;w^{p_2q_1 - p_1q_2}
	- 1
	\Bigg)
	\Bigg(
	\displaystyle\sum_{k_3=0}^{d-1}
	\displaystyle\sum_{k_4=0}^{d-1}
	w^{p_4q_3 - p_3q_4}
	- 1
	\Bigg)
	\nonumber\\
	&=&
	\Bigg(
	\displaystyle\sum_{k_1=0}^{d-1}
	\displaystyle\sum_{k_2=0}^{d-1}
	w^{l_2k_1 - l_1k_2}
	\;w^{k_2q_1 - k_1q_2}
	- 1
	\Bigg)
	\Bigg(
	\displaystyle\sum_{k_3=0}^{d-1}
	\displaystyle\sum_{k_4=0}^{d-1}
	w^{q_4k_3 - q_3k_4}
	- 1
	\Bigg)
	\nonumber\\
	&=&
	\Bigg(
	\displaystyle\sum_{k_1=0}^{d-1}
	\displaystyle\sum_{k_2=0}^{d-1}
	w^{l_2k_1 - l_1k_2}
	\;w^{k_2q_1 - k_1q_2}
	- 1
	\Bigg)(-1).
	\label{eq:M11M2}
\end{eqnarray}

\noindent\rule{\linewidth}{0.4pt}

\begin{eqnarray}
	(M^2M^{1.2})_{ij} &=&
	\Bigg(
	\displaystyle\sum_{k_1=0}^{d-1}
	\displaystyle\sum_{k_2=0}^{d-1}
	w^{l_2k_1 - l_1k_2}
	- 1
	\Bigg)
	\Bigg(
	\displaystyle\sum_{k_3=0}^{d-1}
	\displaystyle\sum_{k_4=0}^{d-1}
	w^{l_4k_3 - l_3k_4}
	w^{p_4q_3 - p_3q_4}
	- 1
	\Bigg)
	\nonumber\\
	&=&
	\Bigg(
	\displaystyle\sum_{k_1=0}^{d-1}
	\displaystyle\sum_{k_2=0}^{d-1}
	w^{l_2k_1 - l_1k_2}
	- 1
	\Bigg)
	\Bigg(
	\displaystyle\sum_{k_3=0}^{d-1}
	\displaystyle\sum_{k_4=0}^{d-1}
	w^{l_4k_3 - l_3k_4}
	w^{k_4q_3 - k_3q_4}
	- 1
	\Bigg)
	\nonumber\\
	&=&
	(-1)
	\Bigg(
	\displaystyle\sum_{k_3=0}^{d-1}
	\displaystyle\sum_{k_4=0}^{d-1}
	w^{l_4k_3 - l_3k_4}
	w^{k_4q_3 - k_3q_4}
	- 1
	\Bigg),
	\label{eq:M2M12}
	\\[2ex]
	(M^{1.2}M^2)_{ij} &=&
	\Bigg(
	\displaystyle\sum_{k_1=0}^{d-1}
	\displaystyle\sum_{k_2=0}^{d-1}
	w^{p_2q_1 - p_1q_2}
	- 1
	\Bigg)
	\Bigg(
	\displaystyle\sum_{k_3=0}^{d-1}
	\displaystyle\sum_{k_4=0}^{d-1}
	w^{l_4k_3 - l_3k_4}
	w^{p_4q_3 - p_3q_4}
	- 1
	\Bigg)
	\nonumber\\
	&=&
	\Bigg(
	\displaystyle\sum_{k_1=0}^{d-1}
	\displaystyle\sum_{k_2=0}^{d-1}
	w^{k_2q_1 - k_1q_2}
	- 1
	\Bigg)
	\Bigg(
	\displaystyle\sum_{k_3=0}^{d-1}
	\displaystyle\sum_{k_4=0}^{d-1}
	w^{l_4k_3 - l_3k_4}
	w^{k_4q_3 - k_3q_4}
	- 1
	\Bigg)
	\nonumber\\
	&=&
	(-1)
	\Bigg(
	\displaystyle\sum_{k_3=0}^{d-1}
	\displaystyle\sum_{k_4=0}^{d-1}
	w^{l_4k_3 - l_3k_4}
	w^{k_4q_3 - k_3q_4}
	- 1
	\Bigg).
	\label{eq:M12M2}
\end{eqnarray}

\noindent\rule{\linewidth}{0.4pt}

\begin{eqnarray}
	(M^{1.1}M^{1.2})_{ij} &=&
	\Bigg(
	\displaystyle\sum_{k_1=0}^{d-1}
	\displaystyle\sum_{k_2=0}^{d-1}
	w^{l_2k_1 - l_1k_2}
	- 1
	\Bigg)
	\Bigg(
	\displaystyle\sum_{k_3=0}^{d-1}
	\displaystyle\sum_{k_4=0}^{d-1}
	w^{p_4q_3 - p_3q_4}
	- 1
	\Bigg)
	\nonumber\\
	&=&
	\Bigg(
	\displaystyle\sum_{k_1=0}^{d-1}
	\displaystyle\sum_{k_2=0}^{d-1}
	w^{l_2k_1 - l_1k_2}
	- 1
	\Bigg)
	\Bigg(
	\displaystyle\sum_{k_3=0}^{d-1}
	\displaystyle\sum_{k_4=0}^{d-1}
	w^{k_4q_3 - k_3q_4}
	- 1
	\Bigg)
	\nonumber\\
	&=& (-1)(-1) = 1, \quad \forall i,j,
	\label{eq:M11M12}
	\\[2ex]
	(M^{1.2}M^{1.1})_{ij} &=&
	\Bigg(
	\displaystyle\sum_{k_1=0}^{d-1}
	\displaystyle\sum_{k_2=0}^{d-1}
	w^{p_2q_1 - p_1q_2}
	- 1
	\Bigg)
	\Bigg(
	\displaystyle\sum_{k_3=0}^{d-1}
	\displaystyle\sum_{k_4=0}^{d-1}
	w^{l_4k_3 - l_3k_4}
	- 1
	\Bigg)
	\nonumber\\
	&=&
	\Bigg(
	\displaystyle\sum_{k_1=0}^{d-1}
	\displaystyle\sum_{k_2=0}^{d-1}
	w^{q_2k_1 - q_1k_2}
	- 1
	\Bigg)
	\Bigg(
	\displaystyle\sum_{k_3=0}^{d-1}
	\displaystyle\sum_{k_4=0}^{d-1}
	w^{l_4k_3 - l_3k_4}
	- 1
	\Bigg)
	\nonumber\\
	&=& (-1)(-1) = 1, \quad \forall i,j.
	\label{eq:M12M11}
\end{eqnarray}

\subsubsection{Proof of (S.2), (S.3) and (S.4)}

For Eq.~\ref{S2}, see Eq.~(\ref{eq:M2M0}) and Eq.~(\ref{eq:M11M12}); for Eq.~\ref{S3}, see Eq.~(\ref{eq:M11M0}) and Eq.~(\ref{eq:M2M0}); and for Eq.~\ref{S4} the  proof is straightforward.

\subsubsection{(S.5) proof}

Let's first see the $(M^{1.1})^2$ product:
\begin{eqnarray}\label{eq:M11M11}
	(M^{1.1}M^{1.1})_{ij} &=&
	\Bigg(
	\displaystyle\sum_{k_1=0}^{d-1}
	\displaystyle\sum_{k_2=0}^{d-1}
	w^{l_2k_1 - l_1k_2}
	\;w^{p_2q_1 - p_1q_2}
	- 1
	\Bigg)
	\Bigg(
	\displaystyle\sum_{k_3=0}^{d-1}
	\displaystyle\sum_{k_4=0}^{d-1}
	1 - 1
	\Bigg)
	\nonumber\\
	&=&
	\Bigg(
	\displaystyle\sum_{k_1=0}^{d-1}
	\displaystyle\sum_{k_2=0}^{d-1}
	w^{l_2k_1 - l_1k_2}
	\;w^{k_2q_1 - k_1q_2}
	- 1
	\Bigg)
	(d^2 - 1)
	=
	-(d^2 - 1)\,(M^2 M^{1.1})_{ij},
\end{eqnarray}
which we get from Eq.~(\ref{eq:M2M11}). Now, for the $k$-th power
\begin{eqnarray}\label{eq:M11M11k_times}
	(M^{1.1})^k &=& (M^{1.1})^{k-2}(M^{1.1})^2
	= -(d^2 - 1)(M^{1.1})^{k-2}(M^2 M^{1.1})
	= -(d^2 - 1)(M^{1.1})^{k-3}(M^2)(M^{1.1})^2
	\nonumber\\
	&=& (d^2 - 1)^2(M^{1.1})^{k-3}(M^2)^2M^{1.1}
	= \ldots = (-1)^{k-1}(d^2 - 1)^{k-1}(M^2)^{k-1}M^{1.1}.
	\label{eq:M11_power}
\end{eqnarray}

\subsubsection{(S.6) proof}

Let's now see the $(M^{1.2})^2$ product:
\begin{eqnarray}\label{eq:M12M12}
	(M^{1.2}M^{1.2})_{ij} &=&
	\Bigg(
	\displaystyle\sum_{k_1=0}^{d-1}
	\displaystyle\sum_{k_2=0}^{d-1}
	1 - 1
	\Bigg)
	\Bigg(
	\displaystyle\sum_{k_3=0}^{d-1}
	\displaystyle\sum_{k_4=0}^{d-1}
	w^{l_4k_3 - l_3k_4}
	w^{p_4q_3 - p_3q_4}
	- 1
	\Bigg)
	\nonumber\\
	&=&
	(d^2 - 1)
	\Bigg(
	\displaystyle\sum_{k_3=0}^{d-1}
	\displaystyle\sum_{k_4=0}^{d-1}
	w^{l_4k_3 - l_3k_4}
	w^{k_4q_3 - k_3q_4}
	- 1
	\Bigg)
	=
	-(d^2 - 1)\,(M^2 M^{1.2})_{ij},
\end{eqnarray}
which we get from Eq.~(\ref{eq:M2M12}). Now, for the $k$-th power
\begin{eqnarray}\label{eq:M12M12k_times}
	(M^{1.2})^k &=& (M^{1.2})^{k-2}(M^{1.2})^2
	= -(d^2 - 1)(M^{1.2})^{k-2}(M^2 M^{1.2})
	= -(d^2 - 1)(M^{1.2})^{k-3}(M^2)(M^{1.2})^2
	\nonumber\\
	&=& (d^2 - 1)^2(M^{1.2})^{k-3}(M^2)^2 M^{1.2}
	= \ldots = (-1)^{k-1}(d^2 - 1)^{k-1}(M^2)^{k-1}M^{1.2}.
	\label{eq:M12_power}
\end{eqnarray}

\subsection{Sub-matrices eigenvalues}

To determine the eigenvalues of the matrices $M^2$, $M^{1.1}$, and $M^{1.2}$, the strategy is to square these matrices and obtain the squared eigenvalues, since this approach simplifies the calculation. All the squared matrices share the same submatrix structure:
\begin{equation}
	\begin{pmatrix}
		A & B & \cdots & B \\
		B & A & \cdots & B \\
		\vdots & \vdots & \ddots & \vdots \\
		B & B & \cdots & A
	\end{pmatrix},
\end{equation}
where $A$ and $B$ are $(d^2 - 1) \times (d^2 - 1)$ matrices that differ from product to product. This structure arises due to the symmetry of the matrix products. The diagonal elements of $A$ correspond to those cases where $\{l_1, l_2, l_3, l_4\} = \{q_1, q_2, q_3, q_4\}$, while the off-diagonal elements satisfy $\{l_1, l_2\} = \{q_1, q_2\}$ and $\{l_3, l_4\} \neq \{q_3, q_4\}$.   For matrix $B$, the diagonal elements correspond to $\{l_1, l_2\} \neq \{q_1, q_2\}$ and $\{l_3, l_4\} = \{q_3, q_4\}$, whereas for the remaining elements, none of the pairs $l_m$ and $q_m$ coincide.

\subsubsection{$M^0$ eigenvalues}

The eigenvalues of an $n \times n$ matrix filled with ones are $0$ (with multiplicity $n-1$) and $n$. For the matrix $M^0$, the corresponding eigenvalues are therefore $\{0, (d^2 - 1)^2\}$.

\subsubsection{$M^2$ eigenvalues}

\begin{eqnarray}
	(M^2M^2)_{ij} &=&
	\Bigg(
	\displaystyle\sum_{k_1=0}^{d-1}
	\displaystyle\sum_{k_2=0}^{d-1}
	w^{l_2k_1 - l_1k_2} \, w^{p_2q_1 - p_1q_2} - 1
	\Bigg)
	\Bigg(
	\displaystyle\sum_{k_3=0}^{d-1}
	\displaystyle\sum_{k_4=0}^{d-1}
	w^{l_4k_3 - l_3k_4} \, w^{p_4q_3 - p_3q_4} - 1
	\Bigg)
	\nonumber\\[1.5ex]
	&=&
	\Bigg(
	\displaystyle\sum_{k_1=0}^{d-1}
	\displaystyle\sum_{k_2=0}^{d-1}
	w^{l_2k_1 - l_1k_2} \, w^{k_2q_1 - k_1q_2} - 1
	\Bigg)
	\Bigg(
	\displaystyle\sum_{k_3=0}^{d-1}
	\displaystyle\sum_{k_4=0}^{d-1}
	w^{l_4k_3 - l_3k_4} \, w^{k_4q_3 - k_3q_4} - 1
	\Bigg).
	\label{eq:M2M2}
\end{eqnarray}

Substituting these values yields the submatrices $A$ and $B$:
\begin{equation}
	A=
	\begin{pmatrix}
		(d^2-1)^2 & -(d^2-1) & \cdots & -(d^2-1) \\
		-(d^2-1) & (d^2-1)^2 & \cdots & -(d^2-1) \\
		\vdots & \vdots & \ddots & \vdots \\
		-(d^2-1) & -(d^2-1) & \cdots & (d^2-1)^2,
	\end{pmatrix}
\end{equation}
\begin{equation}
	B=
	\begin{pmatrix}
		-(d^2-1) & 1 & \cdots & 1 \\
		1 & -(d^2-1) & \cdots & 1 \\
		\vdots & \vdots & \ddots & \vdots \\
		1 & 1 & \cdots & -(d^2-1)
	\end{pmatrix}.
\end{equation}

Since $A$ and $B$ commute, we will rely on  M. H. Ingraham’s note on determinants~\cite{APX:Ingraham1937}. It states that if a square matrix $C = (C_{ij})$ consists of pairwise commuting square subblocks, then the determinant of the matrix built from $\det(C_{ij})$ coincides with $\det(C)$. The determinant is then
\begin{equation}
	\begin{vmatrix}
		a & b & \cdots & b \\
		b & a & \cdots & b \\
		\vdots & \vdots & \ddots & \vdots \\
		b & b & \cdots & a
	\end{vmatrix}=
	\begin{vmatrix}
		a+(n-1)b & b & \cdots & b \\
		0 & a-b & \cdots & b \\
		\vdots & \vdots & \ddots & \vdots \\
		0 & 0 & \cdots & a-b
	\end{vmatrix}
	=(a-b)^{n-1}(a+(n-1)b).
\end{equation}
Hence,
\begin{equation}
	\text{det}((M^2)^2-\lambda I)=\begin{vmatrix}
		A-\lambda I & B & \cdots & B \\
		B & A-\lambda I & \cdots & B \\
		\vdots & \vdots & \ddots & \vdots \\
		B & B & \cdots & A-\lambda I
	\end{vmatrix}=(A-\lambda I-B)^{n-1}(A-\lambda I+(n-1)B),
\end{equation}
where $n = (d^2 - 1)$. To compute the determinant, we rewrite $A$ and $B$ using $J$ as the all-ones matrix and $I$ as the identity, both of size $d^2 - 1$:
\begin{equation}
	A=d^2(d^2-1)I-(d^2-1)J \quad \& \quad B=-d^2I+J.
\end{equation}
After substitution, we obtain:
\begin{equation}
	\text{det}((M^2)^2-\lambda I)=((d^4-\lambda)I-d^2J)^{n-1}((d^2-\lambda)I-J).
\end{equation}
The eigenvalues then satisfy:
\begin{equation}
	|\text{det}((M^2)^2-\lambda I)|=|((d^4-\lambda)I-d^2J)|^{n-1}|((d^2-\lambda)I-J)|=0.
\end{equation}

Because the determinant equals the product of the eigenvalues, the equation reduces to:
\begin{equation}
	((d^4-\lambda)^{n-1}(d^2-\lambda))^{n-1}(d^2-\lambda)^{n-1}(1-\lambda)=0.
\end{equation}
Thus, the possible eigenvalues of $M^2$ are $\{\pm 1, \pm d, \pm d^2\}$.

\subsubsection{$M^{1.1}$ eigenvalues}

\begin{eqnarray}
	(M^{1.1}M^{1.1})_{ij} &=&
	\Bigg(
	\displaystyle\sum_{k_1=0}^{d-1}
	\displaystyle\sum_{k_2=0}^{d-1}
	w^{l_2k_1 - l_1k_2} \, w^{p_2q_1 - p_1q_2} - 1
	\Bigg)
	\Bigg(
	\displaystyle\sum_{k_3=0}^{d-1}
	\displaystyle\sum_{k_4=0}^{d-1}
	1 - 1
	\Bigg)
	\nonumber\\[1.5ex]
	&=&
	\Bigg(
	\displaystyle\sum_{k_1=0}^{d-1}
	\displaystyle\sum_{k_2=0}^{d-1}
	w^{l_2k_1 - l_1k_2} \, w^{k_2q_1 - k_1q_2} - 1
	\Bigg)
	(d^2 - 1)
	\label{eq:M11M11}.
\end{eqnarray}

By substituting the corresponding values, we obtain the matrices $A$ and $B$:
\begin{equation}
	A= (d^2-1)^2 J \quad \& \quad B=-(d^2-1)J.
\end{equation}
Since they commute, the same reasoning applies:
\begin{equation}
	\text{det}((M^{1.1})^2-\lambda I)=(d^2(d^2-1)J-\lambda I)^{n-1}((d^2-1)J-\lambda I),
\end{equation}
where $n = (d^2 - 1)$. Solving for the eigenvalues gives:
\begin{equation}
	|\text{det}((M^{1.1})^2-\lambda I)|=|(d^2(d^2-1)J-\lambda I)|^{n-1}
	|((d^2-1)J-\lambda I)|=0.
\end{equation}
Because the determinant equals the product of the eigenvalues, we find:
\begin{equation}
	((\lambda)^{n-1}(d^2(d^2-1)^2-\lambda))^{n-1}(\lambda)^{n-1}((d^2-1)^2-\lambda)=0.
\end{equation}
Therefore, the eigenvalues of $M^{1.1}$ are $\{0, \pm (d^2 - 1), \pm d(d^2 - 1)\}$.

\subsubsection{$M^{1.2}$ eigenvalues}

\begin{eqnarray}
	(M^{1.2}M^{1.2})_{ij} &=&
	\Bigg(
	\displaystyle\sum_{k_1=0}^{d-1}
	\displaystyle\sum_{k_2=0}^{d-1}
	1 - 1
	\Bigg)
	\Bigg(
	\displaystyle\sum_{k_3=0}^{d-1}
	\displaystyle\sum_{k_4=0}^{d-1}
	w^{l_4k_3 - l_3k_4} \, w^{p_4q_3 - p_3q_4} - 1
	\Bigg)
	\nonumber\\[1.5ex]
	&=&
	(d^2 - 1)
	\Bigg(
	\displaystyle\sum_{k_3=0}^{d-1}
	\displaystyle\sum_{k_4=0}^{d-1}
	w^{l_4k_3 - l_3k_4} \, w^{k_4q_3 - k_3q_4} - 1
	\Bigg).
	\label{eq:M12M12}
\end{eqnarray}

Substituting the relevant parameters yields the matrices $A$ and $B$:
\begin{equation}
	A= B=d^2(d^2-1)I-(d^2-1)J,
\end{equation}
\begin{equation}
	\text{det}((M^{1.2})^2-\lambda I)=(-\lambda I)^{n-1}(d^2(d^2-1)^2I-(d^2-1)^2J-\lambda I),
\end{equation}
\begin{equation}
	|\text{det}((M^{1.2})^2-\lambda I)|=|(-\lambda I)^{n-1}|^{n-1}
	|(d^2(d^2-1)^2I-(d^2-1)^2J-\lambda I)|=0,
\end{equation}
\begin{equation}
	((\lambda)^{n-1})^{n-1}(d^2(d^2-1)^2-\lambda)^{n-1}((d^2-1)^2-\lambda)=0.
\end{equation}
The resulting eigenvalues of $M^{1.2}$ are $\{0, \pm (d^2 - 1), \pm d(d^2 - 1)\}$.